\pdfoutput=1
\documentclass[twocolumn,journal]{IEEEtran}
\usepackage[T1]{fontenc}
\usepackage[latin9]{inputenc}
\usepackage{amsmath}
\usepackage{graphicx}
\usepackage[unicode=true,
 bookmarks=true,bookmarksnumbered=true,bookmarksopen=true,bookmarksopenlevel=1,
 breaklinks=false,pdfborder={0 0 0},pdfborderstyle={},backref=false,colorlinks=false]
 {hyperref}
\hypersetup{pdftitle={Your Title},
 pdfauthor={Your Name},
 pdfpagelayout=OneColumn, pdfnewwindow=true, pdfstartview=XYZ, plainpages=false}

\makeatletter
\usepackage[caption=false,font=footnotesize]{subfig}

\makeatother

\begin{document}
\title{Closed-Loop Wireless Power Transfer with Adaptive Waveform and Beamforming:
\\
Design, Prototype, and Experiment}
\author{Shanpu~Shen,~\IEEEmembership{Member,~IEEE,} Junghoon~Kim,~\IEEEmembership{Member,~IEEE,}
and~Bruno~Clerckx,~\IEEEmembership{Senior~Member,~IEEE}\thanks{Manuscript received; This work was supported in part by the EPSRC
of U.K. under Grant EP/P003885/1 and EP/R511547/1. \textit{(Corresponding
author: Shanpu Shen.)}}\thanks{The authors are with the Department of Electrical and Electronic Engineering,
Imperial College London, London SW7 2AZ, U.K. (e-mail: s.shen@imperial.ac.uk;
junghoon.kim15@imperial.ac.uk; b.clerckx@imperial.ac.uk).}}
\maketitle
\begin{abstract}
In this paper, we design, prototype, and experiment a closed-loop
radiative wireless power transfer (WPT) system with adaptive waveform
and beamforming using limited feedback. Spatial and frequency domains
are exploited by jointly utilizing multi-sine waveform and multi-antenna
beamforming at the transmitter in WPT system to adapt to the multipath
fading channel and boost the output dc power. A closed-loop architecture
based on a codebook design and a low complexity over-the-air limited
feedback using an IEEE 802.15.4 RF interface is proposed. The codebook
consists of multiple codewords where each codeword represents particular
waveform and beamforming. The transmitter sweeps through the codebook
and then the receiver feeds back the index of the optimal codeword,
so that the waveform and beamforming can be adapted to the multipath
fading channel to maximize the output dc power without requiring explicit
channel estimation and the knowledge of accurate Channel State Information.
The proposed closed-loop WPT with adaptive waveform and beamforming
using limited feedback is prototyped using a Software Defined Radio
equipment and measured in a real indoor environment. The measurement
results show that the proposed closed-loop WPT with adaptive waveform
and beamforming can increase the output dc power by up to 14.7 dB
compared with the conventional single-tone and single-antenna WPT
system.
\end{abstract}


\section{Introduction}

\IEEEPARstart{W}{ireless} power transfer (WPT) via radio-frequency
(RF) has gained increasing interests as a promising technology to
energize a large number of low-power and low duty-cycle devices in
applications such as the Wireless Sensor Networks and Internet of
Things \cite{2013_MM_WPT_CUT}. WPT utilizes a dedicated transmitter
to radiate RF energy through a wireless channel and a rectifying antenna
(rectenna) to receive and convert the RF energy into dc power. Compared
with batteries which required to be replaced and recharged periodically,
WPT is more reliable, controllable, user-friendly, and cost-effective.

However, a crucial challenge of radiative WPT is to increase the output
dc power of the rectenna given a fixed transmit power, or equivalently
to enhance end-to-end power transfer efficiency. To that end, the
vast majority of the research has focused on designing efficient rectenna
to increase the output dc power. Various techniques to enhance the
rectenna design include using multiband rectenna \cite{ShanpuShen_2020_TIE_CSong}\nocite{ShanpuShen2017_AWPL_DPTB}\nocite{2018_TMTT_RFEH_OnPaper}-\cite{ShanpuShen2019_TIE_HybridCombining},
multiport rectenna \cite{ShanpuShen2017_TAP_EHPIXEL}\nocite{2019_TMTT_EH_GridArray}\nocite{2019TAP_EH_optimalAngular}\nocite{ShanpuShen2019_TMTT_Freqdepend}-\cite{ShanpuShen_2020_IoTJ_AmRFEH},
compact rectenna \cite{2019TAP_WPT_WeiLinDrivenLoop}, \cite{2019_TMTT_WPT_compactRectenna},
metasurface rectenna \cite{2020_TMTT_WPT_AmRFEH_metasurface}, \cite{2020_Access_Meta_rectenna},
tightly coupled array rectenna \cite{2018TAP_WPT_TCA}, \cite{2020_TMTT_RFEH_TeeShirt},
multiband and broadband rectifier \cite{2019_TMTT_Rectifier_Single_dual_rectifier},
\cite{2017_TMTT_Rectifier_UWB}, high-efficiency rectifier \cite{2019_TMTT_EH_Rectifier_BoosterRegulator}-\nocite{2019_TMTT_Rectifier_Insensitive}\nocite{2020_TMTT_Rectifier_CoupleTransmission}\cite{2020_TMTT_Rectifier_Efficient},
reconfigurable rectifier \cite{ShanpuShen2019_JSSC_Reconfigure},
and hybrid RF-solar harvester \cite{2017_TMTT_RFEH_Solar}, \cite{ShanpuShen2019_TMTT_RF_Solar}.

Another and complementary research area to increase the output dc
power is to design efficient WPT signals \cite{2017_TOC_WPT_YZeng_Bruno_RZhang},
\cite{clerckx2021wireless}. One promising signal strategy is the
waveform. Efficient waveforms can indeed increase the output dc power
in WPT. Due to the nonlinearity of rectifier, the RF-to-dc efficiency
depends on not only the input RF power level but also the waveform
shape \cite{2015_MM_WPT_waveform}. It has been shown through simulations
and experiments that the RF-to-dc efficiency can be improved by waveforms
with high peak to average power ratio (PAPR), such as multi-sine waveforms
\cite{2014TMTT_WPT_MutliSine_Spatial_Power_Combination}-\nocite{2016_TMTT_RFEH_multitone}\nocite{2017_TMTT_RFID_Multisine}\cite{2017_TMTT_WPT_INVIVO}
and other waveforms including orthogonal frequency division multiplexing
(OFDM), white noise, chaotic \cite{2014_MWCL_WPT_Optimal_Waveform}.
However, there are two limitations in these works \cite{2014TMTT_WPT_MutliSine_Spatial_Power_Combination}-\cite{2014_MWCL_WPT_Optimal_Waveform}.
\textit{First}, they ignore the multipath fading in the wireless channel
between the transmitter and the receiver, though multipath makes the
waveform input into the rectifier at the receiver different from the
waveform transmitted by the transmitter. \textit{Second}, they are
based on an open-loop system with the waveform being static and designed
without considering any Channel State Information (CSI) of multipath
fading channel, which degrades the output dc power. Therefore, a closed-loop
WPT system with waveform adaptive to the multipath fading channel
is needed \cite{2018_MM_WPT_Bruno_1GWPT}.

Using multiple antennas at the transmitter with efficient beamforming
is another effective signal strategy to increase the output dc power
in WPT. By using beamforming, the RF signal transmitted by each antenna
element can be coherently added together at the receiver so as to
increase the output dc power \cite{ShanpuShen_2020_TWC_MIMO_WPT}.
Various WPT systems with beamforming have been prototyped including
using digital beamforming through baseband precoding \cite{2019_TMTT_WPT_Selective_Tracking},
analog beamforming through phased array \cite{2020_TMTT_WPT_PhasedArray},
and time-modulation array \cite{2016_TMTT_WPT_TimeArray}, \cite{2016_MM_WPT_SMART}.
Particularly, a selective and tracking WPT system using backscattering
for feedback has been designed in \cite{2019_TMTT_WPT_Selective_Tracking}
and WPT systems with adaptive beamforming using receive signal strength
indicator feedback have been designed in \cite{2017_IEEEACCESS_WPT_BlindBF}\nocite{2017_TWC_WPT_Prototyping_Zolerta}-\cite{2018_TSP_WPT_Prototyping_RSSI}.
In addition, WPT systems with adaptive beamforming using the second
and third harmonics for feedback have been designed in \cite{2018_MWCL_WPT_3rdHarmonic}\nocite{2019_MWCL_WPT_3rdHarmonic2}-\cite{2020_TMTT_WPT_SecondHarmonic}.
However, the limitation of these works \cite{2019_TMTT_WPT_Selective_Tracking}-\cite{2020_TMTT_WPT_SecondHarmonic}
is that they only focus on using beamforming but did not consider
using efficient waveform design. Therefore, together with the two
aforementioned limitations of the WPT waveform designs \cite{2014TMTT_WPT_MutliSine_Spatial_Power_Combination}-\cite{2014_MWCL_WPT_Optimal_Waveform},
it is found that a closed-loop WPT system with both waveform and beamforming
adaptive to the multipath fading channel is needed.

A unified and systematic theoretical design of closed-loop WPT system
with adaptive waveform and beamforming was first conducted in \cite{2016_TSP_WPT_Bruno_Waveform},
with further notable extensions in \cite{2017_TSP_WPT_Bruno_Yang_Large}-\nocite{2017_AWPL_WPT_Bruno_LowComplexity}\nocite{2018_TWC_WPT_Bruno_HYang_LimitedFeedback}\cite{shen2020joint}.
Simulations in those papers demonstrated the significant benefits
of a systematic design of adaptive waveform and beamforming. However,
results were not demonstrated experimentally. In \cite{2019_Junghoon_Prototyping},
and more recently in \cite{2021_WCL_WPT_RangeExpansion}, the first
experimental WPT system with adaptive waveform and beamforming was
prototyped and experimented, and results demonstrated the significant
benefits in terms of dc power and range expansion of a WPT architecture
relying on channel-adaptive waveform and beamforming. However, those
experimental works used complex channel estimation at the receivers
(based on OFDM channel estimation and pilot transmission, reminiscent
of communication system) and cable feedback mechanisms that do not
lend themselves easily to real-world setup with energy-constrained
devices. The limitation of those experimental works is therefore that
they did not address the challenging problem of energy-efficient and
low complexity CSI acquisition at the transmitter. In addition, a
WPT system with distributed antennas using channel-adaptive antenna
selection and frequency selection has been designed, prototyped, and
experimented in \cite{ShanpuShen_2020_TIE_WPT_DAS}. However, antenna
selection is less efficient compared to beamforming and frequency
selection is less efficient compared to multi-sine waveform, which
limit the output dc power.

In this paper, we design, prototype, and experimentally validate the
first closed-loop WPT system with adaptive waveform and beamforming
using over-the-air limited feedback technique to increase the output
dc power. The contributions of the paper are summarized as follows.

\textit{First}, we propose a closed-loop WPT architecture that exploits
frequency and spatial domains and the nonlinearity of rectifier by
jointly utilizing multi-sine waveform and multi-antenna beamforming
at the transmitter to effectively increase the output dc power. The
architecture uniquely relies on a codebook design and a low complexity
over-the-air limited feedback using an IEEE 802.15.4 RF interface.
The codebook is predefined and consists of multiple codewords where
each codeword represents particular waveform and beamforming. During
a training phase, the transmitter sweeps through the codebook and
the receiver measures the output dc power for each codeword and feeds
back the index of the optimal codeword to the transmitter. Then, the
transmitter can transfer power using the optimal waveform and beamforming
during a WPT phase. The operation is repeated periodically. With the
designed codebook and limited feedback, the channel estimation and
accurate CSI can be avoided and more importantly the multi-sine waveform
and multi-antenna beamforming at the transmitter can be optimized
to adapt to the multipath fading channel in real time.

\textit{Second}, we devise, prototype, and experimentally verify the
proposed closed-loop WPT system with adaptive waveform and beamforming
using limited feedback by leveraging a Software Defined Radio (SDR)
equipment. To the authors\textquoteright{} best knowledge, it is the
first prototype of a closed-loop WPT system with adaptive waveform
and beamforming using over-the-air limited feedback. We measure the
proposed WPT system prototype in a real indoor environment. A closed-loop
WPT system based on cable-feedback and an open-loop WPT system are
also measured as comparison benchmarks. The measurement results show
that using closed-loop adaptive multi-sine waveform and multi-antenna
beamforming can effectively increase the output dc power. Compared
with the conventional 1-tone 1-antenna WPT system, the proposed closed-loop
WPT system with adaptive waveform and beamforming can increase the
output dc power by up to 14.7 dB. In addition, compared with the cable-feedback
based closed-loop WPT system, the proposed closed-loop WPT system
using limited feedback can achieve similar performance while it is
more practical since it does not require knowing the CSI.

\textit{Organization}: Section II describes a closed-loop WPT system
model with limited feedback. Section III provides the closed-loop
WPT system design with adaptive waveform and beamforming. Section
IV provides the experimental setup and measurement results. Section
V concludes the work.

\textit{Notations}: Bold lower letters stands for vectors. A symbol
not in bold font represents a scalar. $\Re\left\{ x\right\} $ and
$\left|x\right|$ refer to the real part and modulus of a complex
scalar $x$, respectively. $\left\Vert \mathbf{x}\right\Vert $, $\mathbf{x}^{T}$,
and $\mathbf{x}^{H}$ refer to the $l_{2}$-norm, transpose, and conjugate
transpose of a vector $\mathbf{x}$, respectively.

\section{Closed-Loop Wireless Power Transfer}

\subsection{System Model}

We consider a multi-sine multi-antenna WPT system. There are $M$
antennas at the transmitter and one antenna at the receiver. A multi-sine
waveform consisting of $N$ tones at angular frequencies $\omega_{1}$,
$\omega_{2}$, ..., $\omega_{N}$ is transmitted. The multi-sine waveform
transmitted by the $m$th transmit antenna is given by 
\begin{equation}
x_{m}\left(t\right)=\Re\left\{ \sum_{n=1}^{N}s_{m,n}e^{j\omega_{n}t}\right\} ,
\end{equation}
where $s_{m,n}$ is a complex weight accounting for the magnitude
and phase of the $n$th tone on the $m$th transmit antenna. We group
$s_{m,n}$ into a vector $\mathbf{s}_{n}=\left[s_{1,n},s_{2,n},\ldots,s_{M,n}\right]^{T}$
which characterizes the beamforming at the $n$th tone. Furthermore,
we group $\mathbf{s}_{n}$ $\forall n$ into a vector $\mathbf{s}=\left[\mathbf{s}_{1}^{T},\mathbf{s}_{2}^{T},\ldots,\mathbf{s}_{N}^{T}\right]^{T}$
which characterizes the waveform and beamforming. The transmitter
is subject to a transmit power constraint given by 
\begin{equation}
\frac{1}{2}\left\Vert \mathbf{s}\right\Vert ^{2}\leq P,
\end{equation}
where $P$ denotes the transmit power.

The multi-sine waveform transmitted by the multiple transmit antennas
propagate through a wireless channel. The waveform received by the
receiver can be expressed as
\begin{align}
y\left(t\right) & =\Re\left\{ \sum_{n=1}^{N}\mathbf{h}_{n}\mathbf{s}_{n}e^{j\omega_{n}t}\right\} ,\label{eq:received waveform}
\end{align}
where $\mathbf{h}_{n}=\left[h_{1,n},h_{2,n},...,h_{M,n}\right]$ with
$h_{m,n}$ referring to the complex channel gain between the $m$th
transmit antenna and the receive antenna at the $n$th angular frequency.
Due to the multipath fading channel, the received waveform $y\left(t\right)$
is different from the transmitted waveform $x_{m}\left(t\right)$.
Therefore, we need to consider the multipath fading channel when designing
the optimal waveform and beamforming for WPT. From \eqref{eq:received waveform},
the received RF power is given by 
\begin{equation}
P_{\mathrm{RF}}=\frac{1}{2}\sum_{n=1}^{N}\left|\mathbf{h}_{n}\mathbf{s}_{n}\right|^{2}.\label{eq:received power}
\end{equation}
The received waveform is input into a rectifier to generate output
dc power. The output dc power is denoted as $P_{\mathrm{DC}}$ and
given by 
\begin{equation}
P_{\mathrm{DC}}=P_{\mathrm{RF}}\eta\left(y\left(t\right)\right)=P_{\mathrm{DC}}\left(\mathbf{h}_{1},\mathbf{h}_{2},...,\mathbf{h}_{N},\mathbf{s}\right),
\end{equation}
where $\eta\left(y\left(t\right)\right)$ refers to the RF-to-dc efficiency
with the input waveform $y\left(t\right)$. From \eqref{eq:received waveform}
and \eqref{eq:received power}, the output dc power is a function
of the wireless channel gains $\mathbf{h}_{n}$ $\forall n$ and the
waveform and beamforming weight vector $\mathbf{s}$, denoted as $P_{\mathrm{DC}}\left(\mathbf{h}_{1},\mathbf{h}_{2},...,\mathbf{h}_{N},\mathbf{s}\right)$.

Assuming the CSI, i.e. the wireless channel gains $\mathbf{h}_{n}$
$\forall n$, is known at the transmitter, the waveform and beamforming
weight vector $\mathbf{s}$ can be optimized to adapt to the channel
to maximize the output dc power, which can be formulated as 
\begin{align}
\underset{\mathbf{s}}{\mathrm{max}}\;\;\; & P_{\mathrm{DC}}\left(\mathbf{h}_{1},\mathbf{h}_{2},...,\mathbf{h}_{N},\mathbf{s}\right)\label{eq:OP2-object}\\
\mathsf{\mathrm{s.t.}}\;\;\,\; & \frac{1}{2}\left\Vert \mathbf{s}\right\Vert ^{2}\leq P.\label{eq:OPT2-constraint-1}
\end{align}
In the next subsections, we provide two strategies for designing adaptive
waveform and beamforming to maximize the output dc power.

\subsection{Scaled Matched Filter}

Scaled Matched Filter (SMF) is a low complexity strategy to optimize
the waveform and beamforming \cite{2017_AWPL_WPT_Bruno_LowComplexity}.
Using SMF, the beamforming weight vector $\mathbf{s}_{n}$ can be
expressed as 
\begin{equation}
\mathbf{s}_{n}=c\left\Vert \mathbf{h}_{n}\right\Vert ^{\beta}\frac{\mathbf{h}_{n}^{H}}{\left\Vert \mathbf{h}_{n}\right\Vert },\:\forall n,
\end{equation}
where $\beta\geq1$ is a parameter controlling the magnitude ($l_{2}$-norm)
of $\mathbf{s}_{n}$ and $c$ is a constant satisfying the transmit
power constraint and is given by 
\begin{equation}
c=\sqrt{\frac{2P}{\sum_{n=1}^{N}\left\Vert \mathbf{h}_{n}\right\Vert ^{2\beta}}}.
\end{equation}
The SMF waveform and beamforming design is only a function of the
single parameter $\beta$. We set $\beta=3$ in this work. In \cite{2017_AWPL_WPT_Bruno_LowComplexity},
it has been shown through simulation that the performance of SMF strategy
with $\beta=3$ is close to that of the optimal waveform and beamforming
design proposed in \cite{2016_TSP_WPT_Bruno_Waveform} but at a much
lower computational complexity.

Although the SMF has good performance and low computational complexity,
it requires that the transmitter has the knowledge of CSI, i.e. $\mathbf{h}_{n}$
$\forall n$. Acquiring the CSI at the transmitter is challenging
in WPT since it needs channel estimation which increases the power
consumption and circuit complexity. To overcome this challenge, in
the next subsection, we provide another waveform and beamforming design
strategy based on limited feedback.

\subsection{Limited Feedback}

\begin{figure}[t]
\begin{centering}
\includegraphics[width=8.8cm]{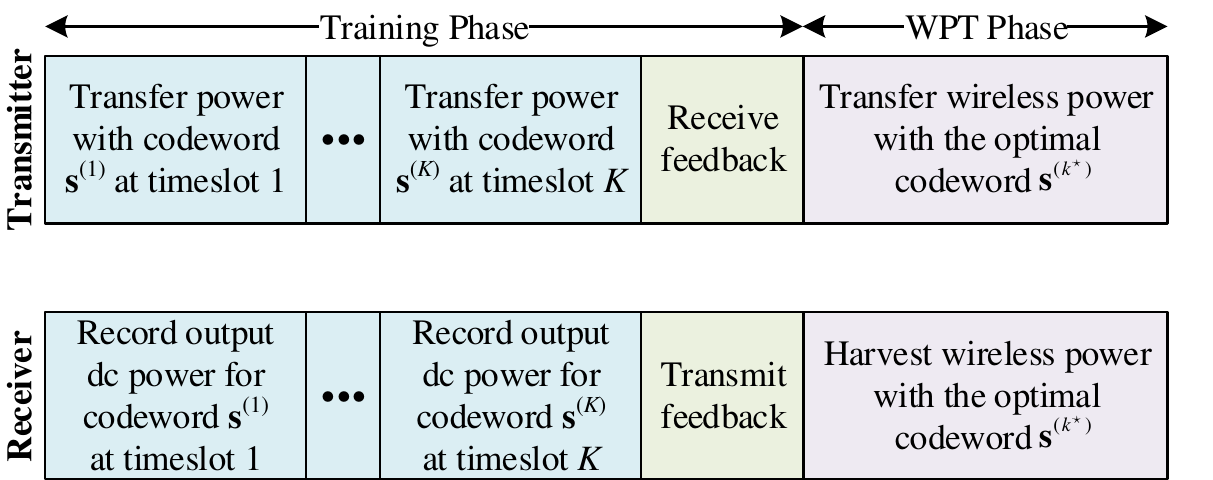}
\par\end{centering}
\caption{\label{fig:Frame}Time frame for the closed-loop WPT using limited
feedback strategy.}
\end{figure}

Consider a codebook consisting of $K$ codewords, denoted as $\mathbf{s}^{\left(1\right)}$,
$\mathbf{s}^{\left(2\right)}$, ..., $\mathbf{s}^{\left(K\right)}$,
with each codeword representing a particular waveform and beamforming
weight vector. The basic idea of limited feedback strategy is to select
the optimal codeword in the codebook to maximize the output dc power.
Specifically, the transmitter and receiver cooperatively work frame
by frame and the frame period is designed to be smaller than the coherence
time of the channel such that the wireless channel gains $\mathbf{h}_{n}$
$\forall n$ are constant during one frame. As shown in Fig. \ref{fig:Frame},
each frame has two phases: training phase and WPT phase. The training
phase is to find the optimal codeword, i.e. the optimal waveform and
beamforming, while the WPT phase is to transfer the wireless power
with the optimal waveform and beamforming. During the training phase,
the transmitter sequentially chooses the $K$ codewords as its waveform
and beamforming to transfer wireless power while in the meantime the
receiver will record the output dc power for each codeword. The output
dc power for the $k$th codeword $\mathbf{s}^{\left(k\right)}$ is
given by $P_{\mathrm{DC}}\left(\mathbf{h}_{1},\mathbf{h}_{2},...,\mathbf{h}_{N},\mathbf{s}^{\left(k\right)}\right)$.
After recording the output dc power for all the codewords, the receiver
can find the index of the optimal codeword maximizing the output dc
power as 
\begin{equation}
k^{\star}=\underset{k=1,\ldots,K}{\mathrm{argmax}}\;P_{\mathrm{DC}}\left(\mathbf{h}_{1},\mathbf{h}_{2},...,\mathbf{h}_{N},\mathbf{s}^{\left(k\right)}\right).\label{eq:maxmax}
\end{equation}
Then, the receiver feeds back the index of the optimal codeword $k^{\star}$
to the transmitter. Since there are $K$ codewords in the codebook,
we only need to feed back $\left\lceil \log_{2}K\right\rceil $ bits
representing the index of the optimal codeword, where $\left\lceil \cdot\right\rceil $
refers to the ceiling function. Once the transmitter receives the
feedback $k^{\star}$, it can use the optimal codeword $\mathbf{s}^{\left(k^{\star}\right)}$
as its waveform and beamforming to efficiently transfer wireless power
during the WPT phase.

In contrast with the SMF strategy, the limited feedback strategy can
adapt to the wireless channel to increase the output dc power without
requiring the knowledge of CSI, i.e. $\mathbf{h}_{n}$ $\forall n$.
Due to such benefit, we focus on using the limited feedback strategy
to implement our closed-loop WPT system in this work. The key of limited
feedback strategy in WPT is the design of an efficient codebook, which
should have diverse codewords to match different wireless channel
gains. In this work, we design an efficient codebook by following
the approach proposed in \cite{2018_TWC_WPT_Bruno_HYang_LimitedFeedback}.
Such approach leverages the statistics of the multipath fading channel
and the rectifier nonlinearity and simulation results have shown that
it can provide high output dc power. In addition, increasing the codebook
size $K$ is beneficial to increase the output dc power as there are
more codewords to select, however, at the expense of feeding back
more bits.

In the next section, we provide a design for the closed-loop WPT system
with adaptive waveform and beamforming using limited feedback.

\section{Closed-Loop WPT System Design}

\begin{figure}[t]
\begin{centering}
\includegraphics[width=8.5cm]{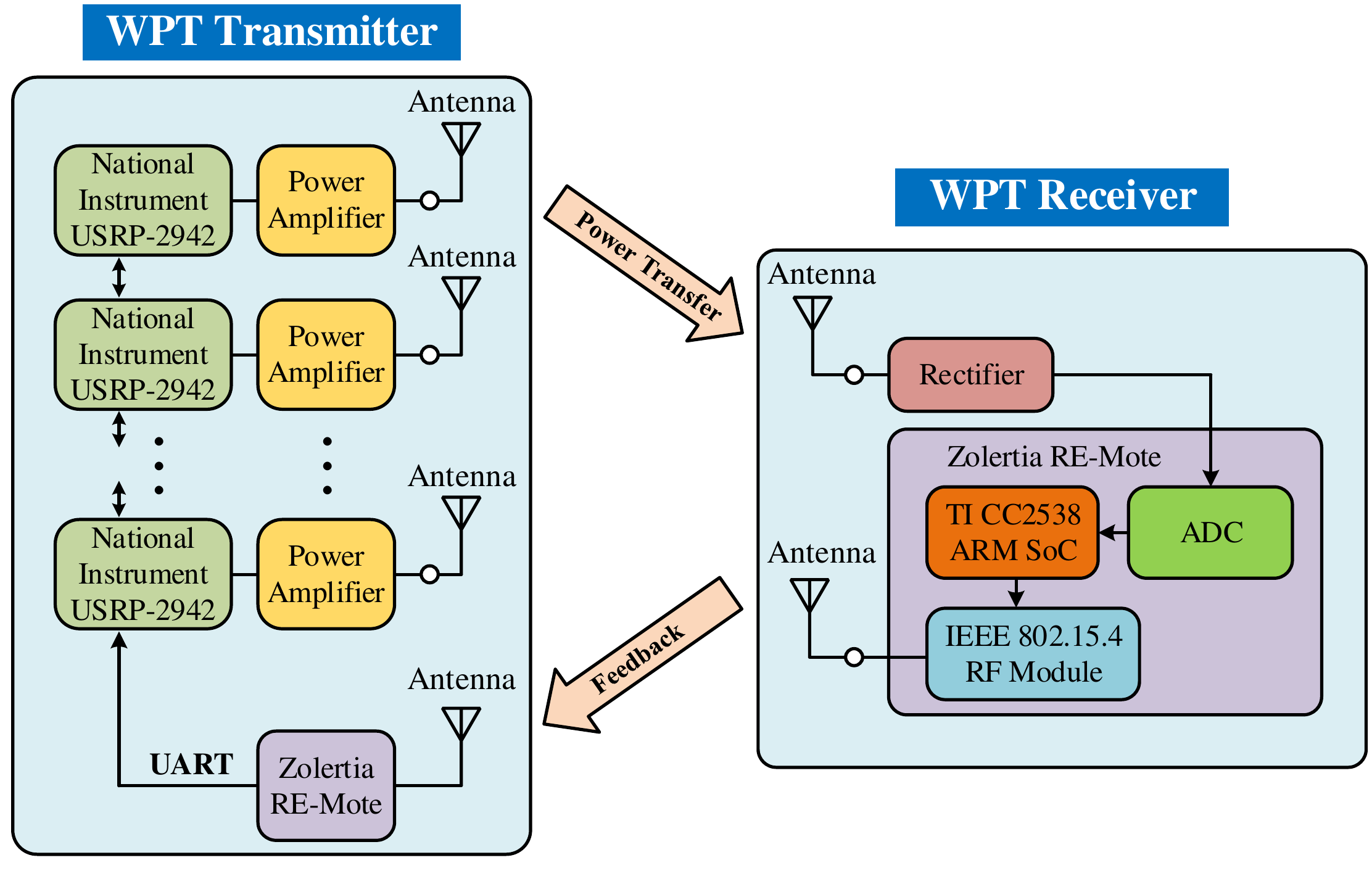}
\par\end{centering}
\caption{\label{fig:Diagram }Schematic diagram of the closed-loop WPT system
with adaptive waveform and beamforming using limited feedback.}
\end{figure}

The schematic diagram of the proposed closed-loop WPT system with
adaptive waveform and beamforming using limited feedback is shown
in Fig. \ref{fig:Diagram }. In the following subsections, we separately
describe the designs for transmitter, receiver, and flow chart of
the closed-loop WPT system.

\subsection{Transmitter Design}

The transmitter consists of two parts. The first part is made up by
multiple antennas, power amplifiers, and SDR equipment, which is used
to generate and radiate RF signals with waveform and beamforming characterized
by different codewords. There are $M$ transmit antennas in the first
part and we consider three cases with $M=1$, 2, and 4. All the antennas
are identical 2.4-GHz monopole antennas which have omnidirectional
radiation patterns with 3 dBi antenna gain and 85\% radiation efficiency.
Each antenna is connected to a power amplifier, Mini-Circuits ZHL-16W-43-S+,
which has a gain of 45 dB and amplifies the RF signal generated by
SDR equipment, and the total transmit power is set to 33 dBm (2W).
The SDR equipment, National Instrument (NI) USRP-2942, can generate
multi-sine waveforms with different magnitudes and phases at each
tone. The multi-sine waveform has $N$ tones and we consider four
cases with $N=1$, 2, 4, and 8. The $N$ tones are centered around
2.4 GHz with a uniform frequency gap $\Delta_{f}=B/N$ where the bandwidth
$B=10$ MHz. We use multiple USRP-2942 to generate multi-sine waveform
for each transmit antenna so as to implement the multi-sine waveform
and multi-antenna beamforming.

\begin{figure}[t]
\begin{centering}
\includegraphics[scale=0.5]{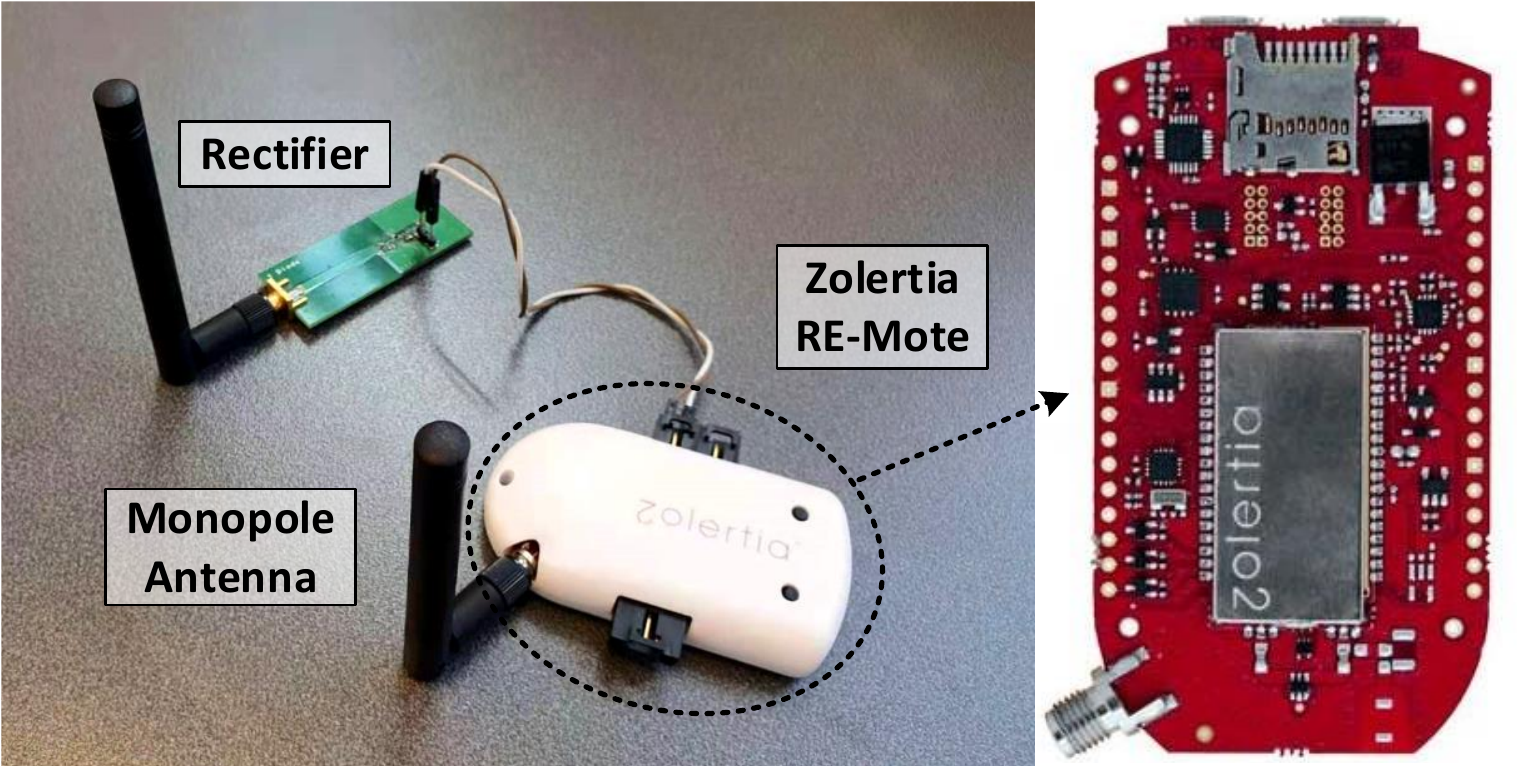}
\par\end{centering}
\caption{\label{fig:receiver}Photo of the receiver in the proposed closed-loop
WPT system and the Zolertia RE-Mote.}
\end{figure}

The second part is made up by a 2.4-GHz monopole antenna and a Zolertia
RE-Mote, which is used to communicate with the receiver and acquire
the optimal codeword fed back from the receiver. The Zolertia RE-Mote
is a hardware development platform consisting of the Texas Instruments
CC2538 ARM Cortex-M3 system on chip (SoC) and an on-board 2.4 GHz
IEEE 802.15.4 RF interface. The photo of the Zolertia RE-Mote is shown
in Fig. \ref{fig:receiver}. In the Zolertia RE-Mote, we use a Contiki
operating system as a software platform. The Zolertia RE-Mote at the
transmitter is used to communicate with the receiver, which is also
equipped with a Zolertia RE-Mote, through the built-in IEEE 802.15.4
RF interface operating at 2.42 GHz (that is different from the frequency
of the multi-sine waveform for WPT). Specifically, the Zolertia RE-Mote
at the transmitter sends a message to the receiver to start the program
and it also receives the index of the optimal codeword fed back from
the receiver, so that the transmitter can transfer the wireless power
with the optimal waveform and beamforming to maximize the output dc
power.

\subsection{Receiver Design}

\begin{figure}[t]
\begin{centering}
\includegraphics[width=8cm]{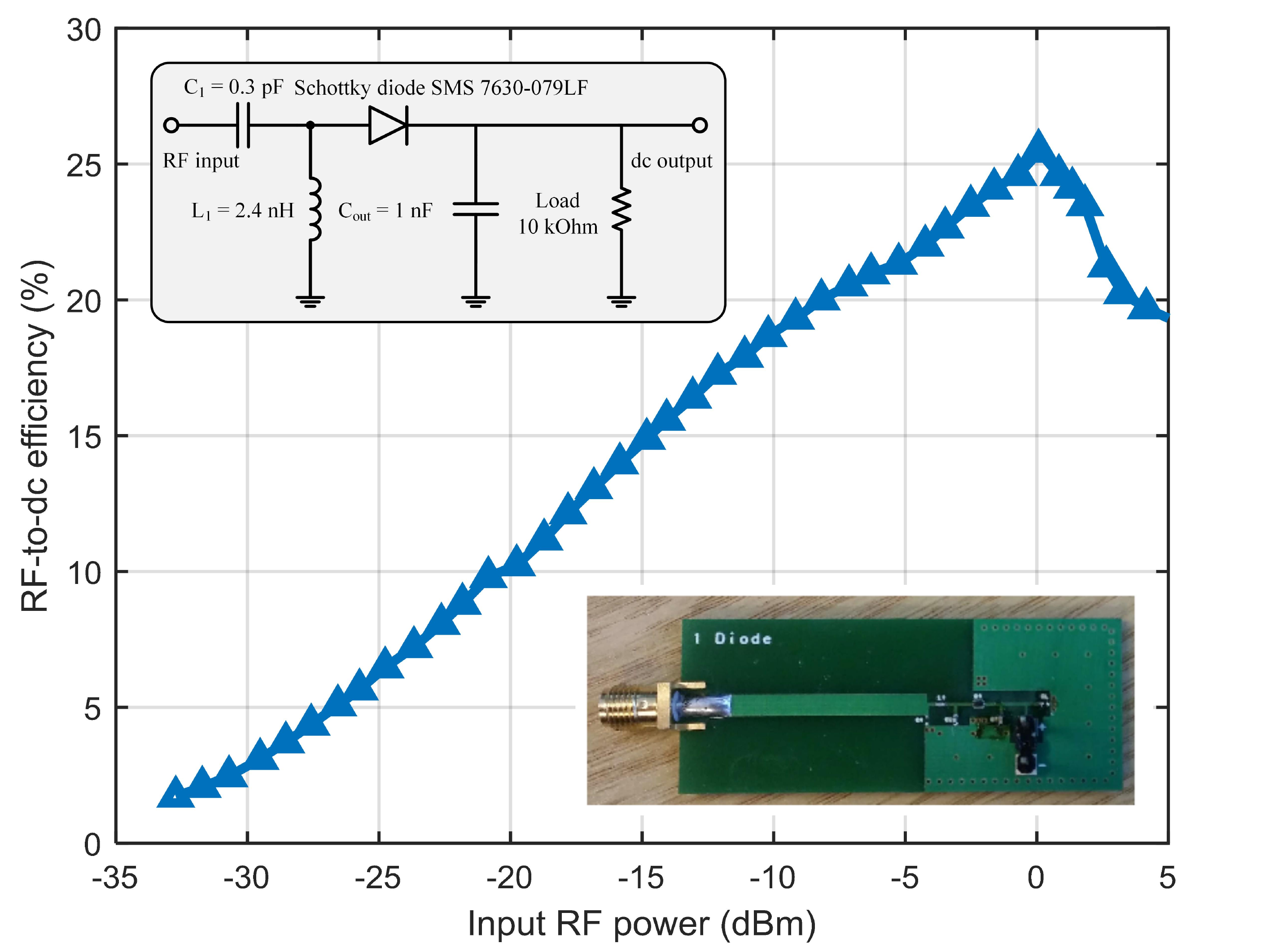}
\par\end{centering}
\caption{\label{fig:rectifier}Topology of the single diode rectifier and its
measured RF-to-dc efficiency.}
\end{figure}

The receiver consists of two parts as shown in Fig. \ref{fig:receiver}.
The first part is a rectenna that receives RF signal and converts
it to dc power. It consists of a single diode rectifier and 2.4-GHz
monopole antenna with 3 dBi gain and 85\% radiation efficiency. We
reuse the single diode rectifier design in \cite{ShanpuShen_2020_TIE_WPT_DAS}
for simplicity because this work focuses on designing a closed-loop
WPT system with adaptive waveform and beamforming instead of focusing
on rectifier design. The topology and measured RF-to-dc efficiency
of the single diode rectifier are shown in Fig. \ref{fig:rectifier}.
The rectifier consists of an impedance matching network, a single
rectifying diode, a low pass filter, and a load. The Schottky diode
Skyworks SMS7630 is chosen as the rectifying diode as it has a low
turn-on voltage, which is suitable for low power rectifier. Common
materials including the 1.6-mm-thick FR-4 substrate and lumped elements
are used to simplify the rectifier fabrication. The values of the
elements in the matching network and low pass filter are optimized
to maximize RF-to-dc efficiency at low input RF power.

The second part is made up by a 2.4-GHz monopole antenna and a Zolertia
RE-Mote, which is used to record the output dc voltage of the rectenna,
communicate with the transmitter, and feed back the index of the optimal
codeword to the transmitter. The Zolertia RE-Mote at the receiver
receives the message from the transmitter to start the program and
then it records the output dc voltage of the rectenna through a built-in
analog-to-digital converter (ADC). The recorded output dc voltages
are processed by the CC2538 ARM Cortex-M3 SoC in the Zolertia RE-Mote
and the index of the optimal codeword is found and fed back to the
transmitter through the built-in IEEE 802.15.4 RF interface so that
the transmitter can transfer the wireless power with the optimal waveform
and beamforming.

\subsection{Flow Chart}

\begin{figure}[t]
\begin{centering}
\includegraphics[width=8.5cm]{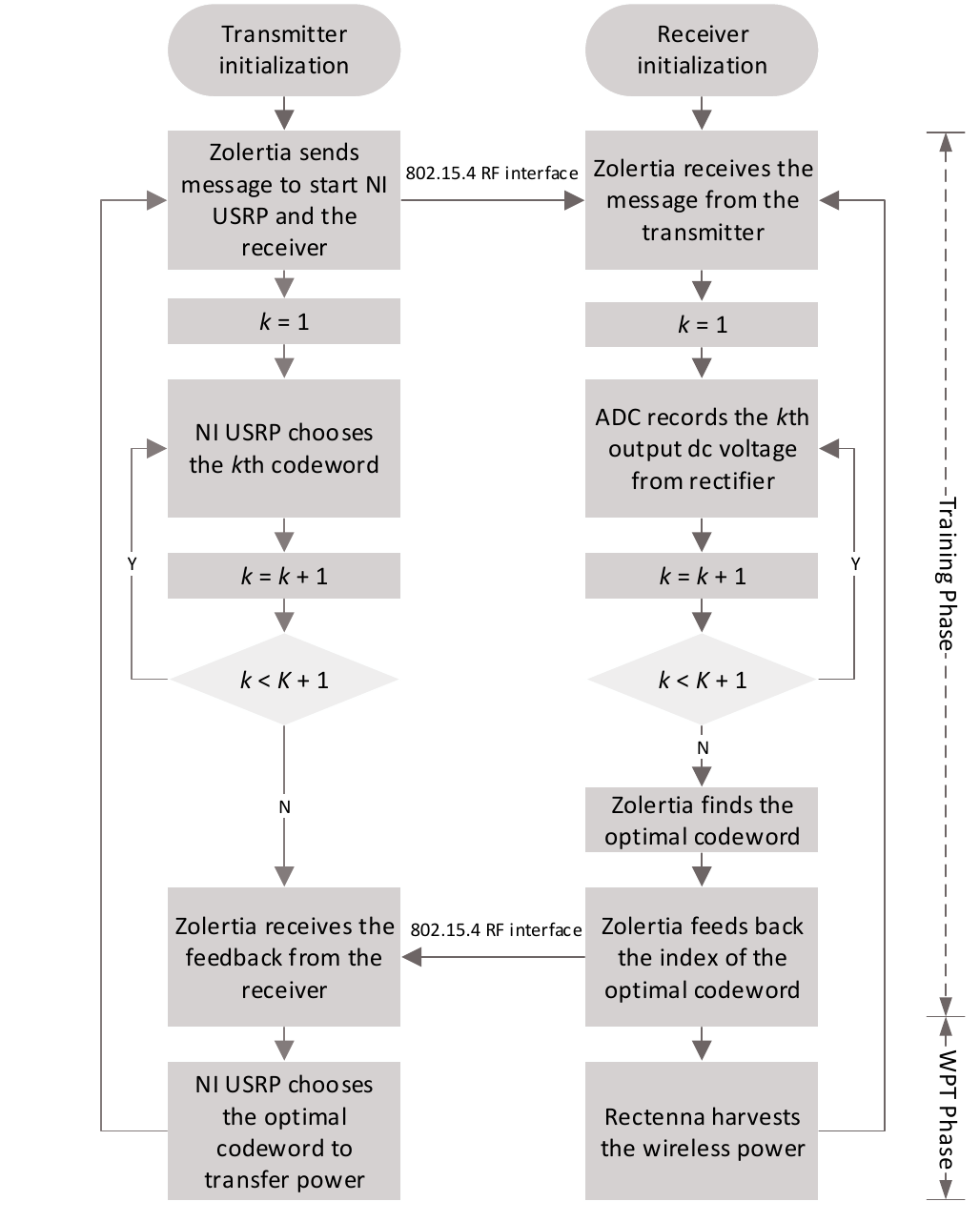}
\par\end{centering}
\caption{\label{fig:flow chart}Flow chart of the closed-loop WPT system with
adaptive waveform and beamforming using limited feedback.}
\end{figure}

The flow chart of the adaptive closed-loop WPT system with adaptive
waveform and beamforming using limited feedback is shown in Fig. \ref{fig:flow chart}.
The transmitter and receiver cooperatively work frame by frame. Each
frame has two phases: training phase and WPT phase, as shown in Fig.
\ref{fig:flow chart}. The training phase is to find the optimal codeword,
i.e. the optimal waveform and beamforming, while the WPT phase is
to transfer the wireless power with the optimal waveform and beamforming.

In the training phase, the Zolertia RE-Mote at the transmitter first
sends a message to the receiver through the built-in IEEE 802.15.4
RF interface so that the receiver will start to work. It also sends
a message to the NI USRP-2942 through UART so that the NI USRP-2942
will start to work. The NI USRP-2942 sequentially chooses each codeword
as its waveform and beamforming to transfer wireless power. The time
duration for transferring wireless power for each codeword is $T_{s}=10\:\mathrm{ms}$.
In the meantime, the receiver will measure and record the corresponding
output dc voltage of the rectenna for each codeword through the built-in
ADC in Zolertia RE-Mote. After the Zolertia RE-Mote at the receiver
records the output dc voltage for all codewords, it finds the optimal
codeword that maximizes the output dc voltage. Generally, for a codebook
having $K$ codewords, the receiver only needs to feed back $\left\lceil \log_{2}K\right\rceil $
bits representing the index of the optimal codeword. In this work,
we consider codebooks having 2, 4, 8, 16, 32, and 64 codewords so
that the receiver only needs to feed back 1, 2, 3, 4, 5, and 6 bits,
respectively, to the transmitter through the IEEE 802.15.4 RF interface.
The Zolertia RE-Mote at the transmitter forwards the received feedback
to NI USRP-2942 through UART so that the transmitter can choose the
optimal codewords as its waveform and beamforming to transfer wireless
power. By this way, we can implement the limited feedback closed-loop
WPT system where the waveform and beamforming are adaptive to the
wireless fading channel to increase the output dc power without requiring
the knowledge of CSI. The time duration for the training phase is
$KT_{s}$. $T_{s}$ is dependent on the clock and timer setup in NI
USRP-2942, which can be modified by programming. To accelerate the
training phase, a smaller $T_{s}$ can be set in NI USRP-2942, however,
$T_{s}$ cannot be too small since the output dc voltage for a given
codeword needs some time to be stable for ADC sampling. If $T_{s}$
is very small, the output dc voltage is not stable and the dc voltage
sampled by ADC is not accurate so that the optimal codeword cannot
be found.

In the WPT phase, the transmitter transfers the wireless power with
the optimal waveform and beamforming. In the meantime, the receiver
harvests the wireless power. When the WPT phase is over, it goes to
the next frame and the time duration for one frame is set as $T=KT_{s}+T_{p}=2\:\mathrm{s}$
where $T_{p}$ denotes the time duration for the WPT phase. In other
words, the proposed closed-loop WPT system periodically adapts to
the wireless fading channel to maximize the output dc power.

It is should be noted that the proposed closed-loop WPT system inherently
captures the nonlinearity of rectifier since the selection of the
codeword is made at the output dc power level (instead of RF power),
hence capturing the influence of the input signal on the RF-to-dc
efficiency.

\section{Closed-Loop WPT System Experiment}

\begin{figure}[t]
\begin{centering}
\includegraphics[width=8.9cm]{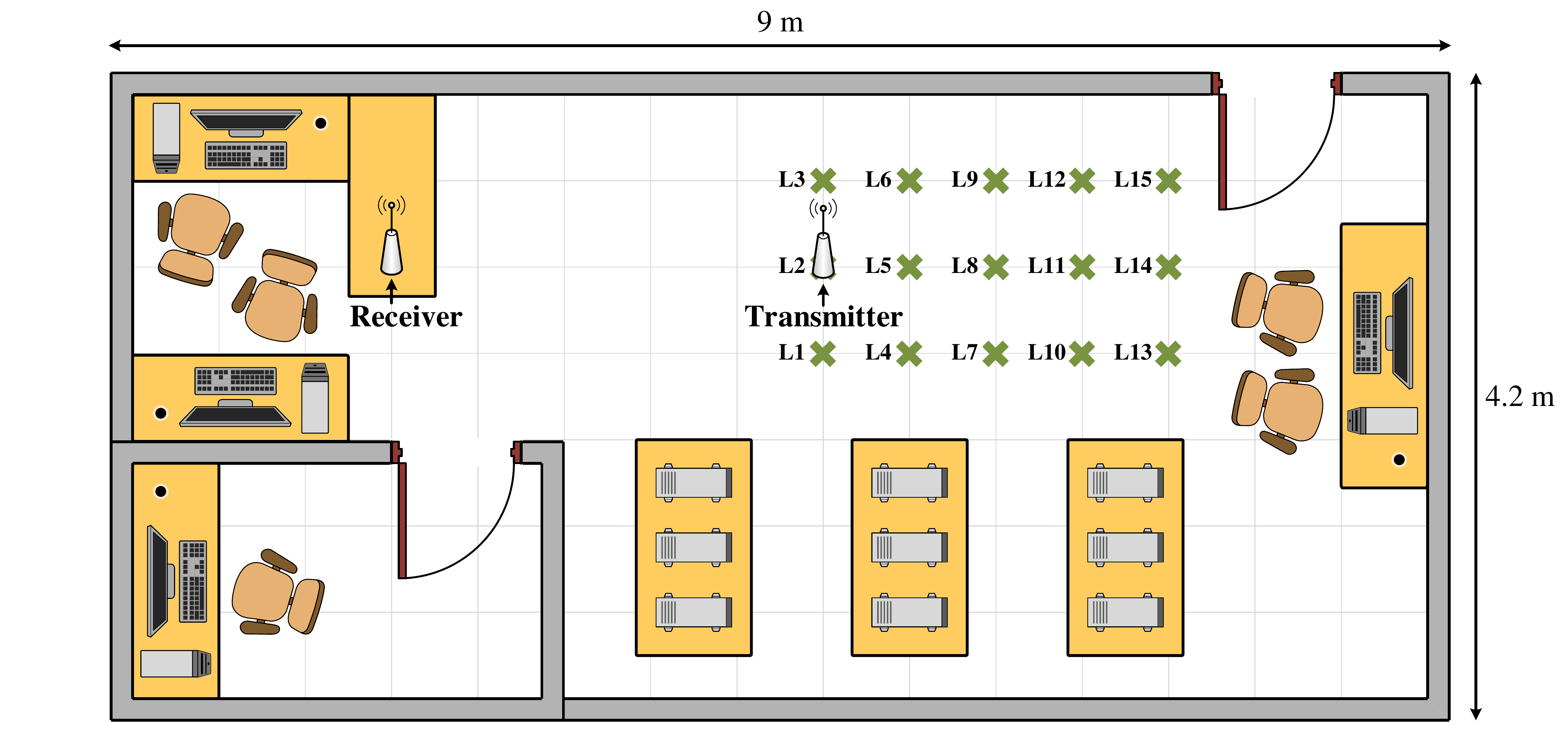}
\par\end{centering}
\caption{\label{fig:map}Illustration of the indoor environment for measurement.}
\end{figure}

\begin{figure}[t]
\begin{centering}
\includegraphics[width=8.6cm]{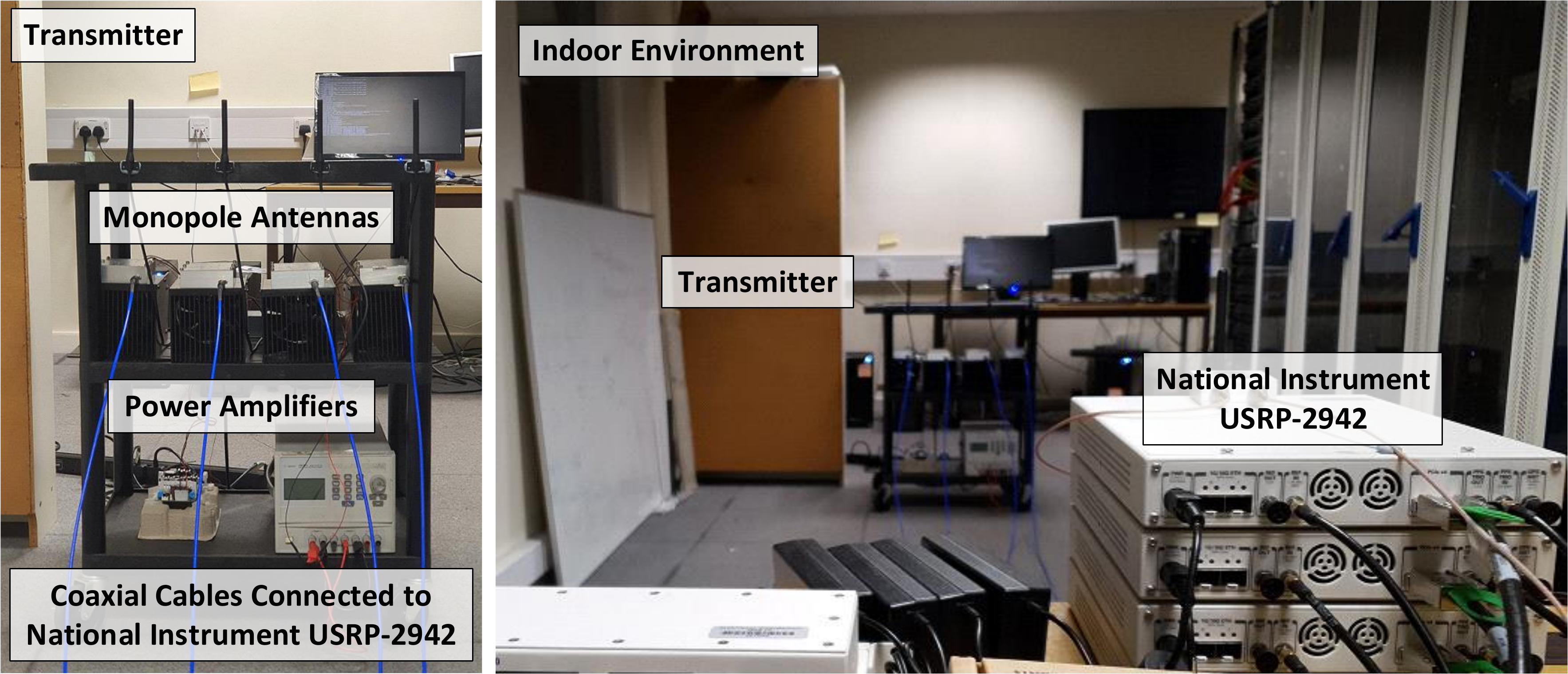}
\par\end{centering}
\caption{\label{fig:measurement}Photos of the proposed closed-loop WPT system
measurement in an indoor environment.}
\end{figure}

To verify the proposed closed-loop WPT system with adaptive waveform
and beamforming using limited feedback, we prototype and experiment
it in a $4.2\mathrm{m}\times9\mathrm{m}$ indoor environment. As illustrated
in Fig. \ref{fig:map}, the indoor environment is equipped with common
facilities such as chairs, tables, and computers, so that multipath
fading exists in the wireless channel. In Fig. \ref{fig:map}, the
transmitter is placed at $3\times5$ different locations marked as
L1, L2, ..., and L15 while the receiver is placed at a fixed location,
so as to measure the performance of the proposed closed-loop WPT system
at different locations. The photos of the proposed closed-loop WPT
system measurement in an indoor environment are shown in Fig. \ref{fig:measurement}.

We consider using codebooks having 2, 4, 8, 16, 32, and 64 codewords
for the proposed closed-loop WPT system, which correspond to feeding
back 1, 2, 3, 4, 5, and 6 bits. In addition, we also consider two
benchmarks for comparison. The first benchmark is the SMF, which is
a closed-loop strategy as shown in Section II. For the SMF strategy,
the receiver performs complex channel estimation by using OFDM channel
estimation and pilot transmission as shown in \cite{2019_Junghoon_Prototyping}
and then feeds back the estimated CSI to the transmitter using cable.
Such cable-based feedback is not practical but herein we just use
it as a benchmark since it provides almost perfect CSI feedback (equivalently
a large number of bits of feedback) to the transmitter. On the other
hand, the second benchmark is the uniform power allocation (UP), which
is an open-loop strategy allocating the same magnitude and phase to
multiple tones and multiple antennas at the transmitter. Specifically,
the complex weights for the UP strategy are given by 
\begin{equation}
s_{m,n}=\sqrt{\frac{2P}{MN}},\forall m,\:n,
\end{equation}
which are independent of the CSI. Therefore, the UP strategy can be
equivalently viewed as 0 bit of feedback.

\subsection{Adaptive Beamforming Only}

\begin{figure}[t]
\begin{centering}
\includegraphics[width=8.5cm]{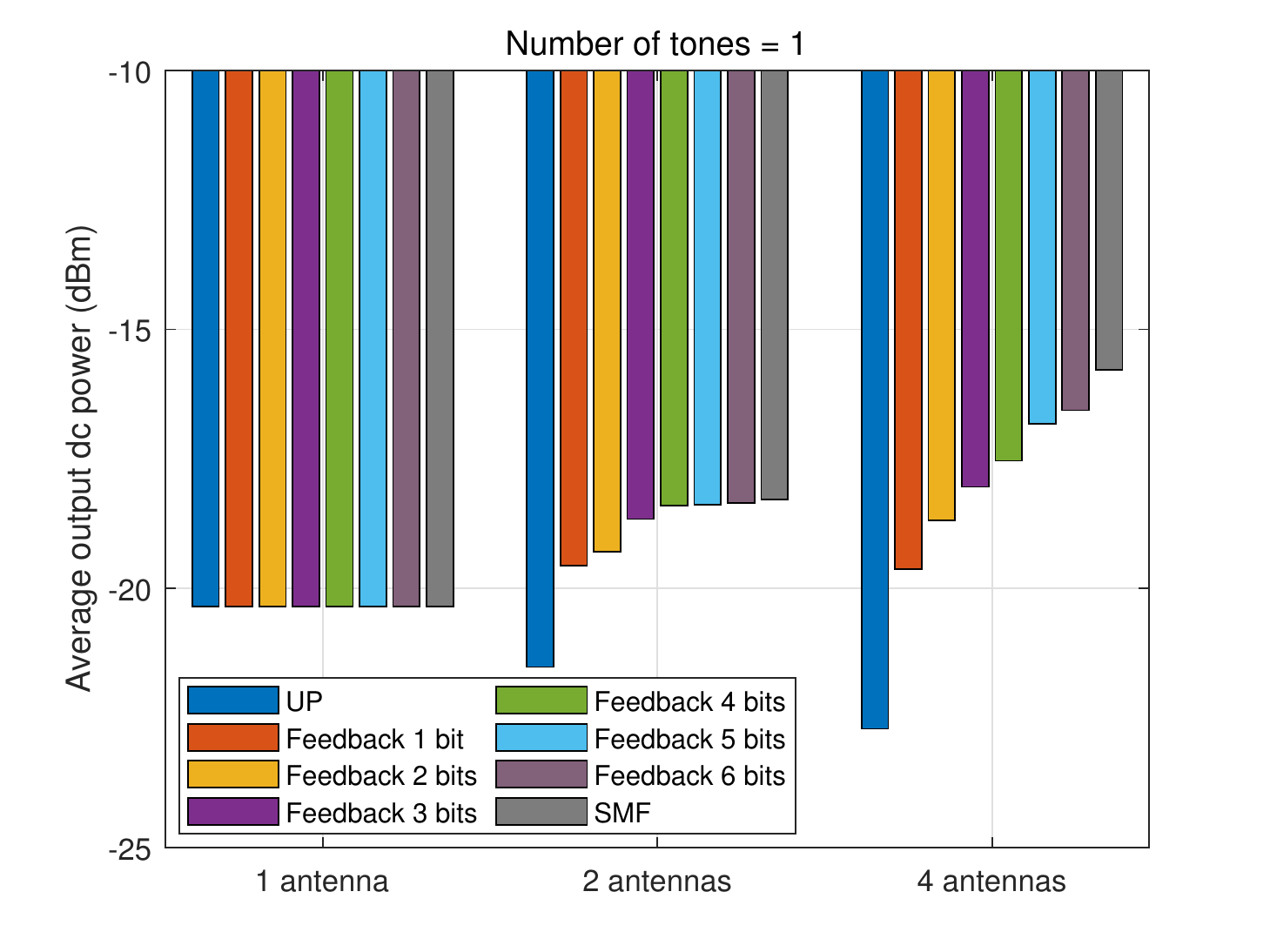}
\par\end{centering}
\caption{\label{fig:BF}Average output dc power for the three strategies versus
the number of transmit antennas with only 1 tone.}
\end{figure}

First, we show the benefit of closed-loop WPT system with only adaptive
beamforming using limited feedback. To that end, we consider the three
strategies with only 1 tone and different numbers of transmit antennas,
1, 2, and 4. We use a multimeter to measure the output dc power of
the rectifier at locations L1-L15 and average the output dc power
over the 15 locations. The average output dc power for the three strategies
versus the number of transmit antennas with only 1 tone is shown in
Fig. \ref{fig:BF}. We make the following observations.

1) The average output dc power increases with the number of transmit
antennas for the SMF and limited feedback strategies, showing the
benefit of multi-antenna adaptive beamforming to increase the output
dc power. For the open-loop UP strategies, increasing the number of
transmit antenna however reduces the output dc power. It is because
the beam of multiple antennas using the UP strategy has fixed direction
and narrow beamwidth, which leads to beam misalignment between the
transmitter and receiver and reduces the output dc power.

2) The closed-loop SMF and limited feedback strategies achieves higher
output dc power than the open-loop strategy UP, showing the benefit
of using closed-loop strategies to adapt beamforming to the wireless
multipath fading channels.

3) The average output dc power for the limited feedback strategy increases
with the number of feedback bits. In other words, using more codewords
is beneficial to increase the output dc power. Particularly, when
the codebook has enough diverse codewords, the limited feedback strategy
can achieve a similar performance to the SMF strategy while having
the benefit that it does not require knowing the CSI.

Overall, the measurement results in Fig. \ref{fig:BF} show the benefit
of closed-loop WPT system with only adaptive beamforming using limited
feedback to increase the output dc power. 

\subsection{Adaptive Waveform Only}

\begin{figure}[t]
\begin{centering}
\includegraphics[width=8.5cm]{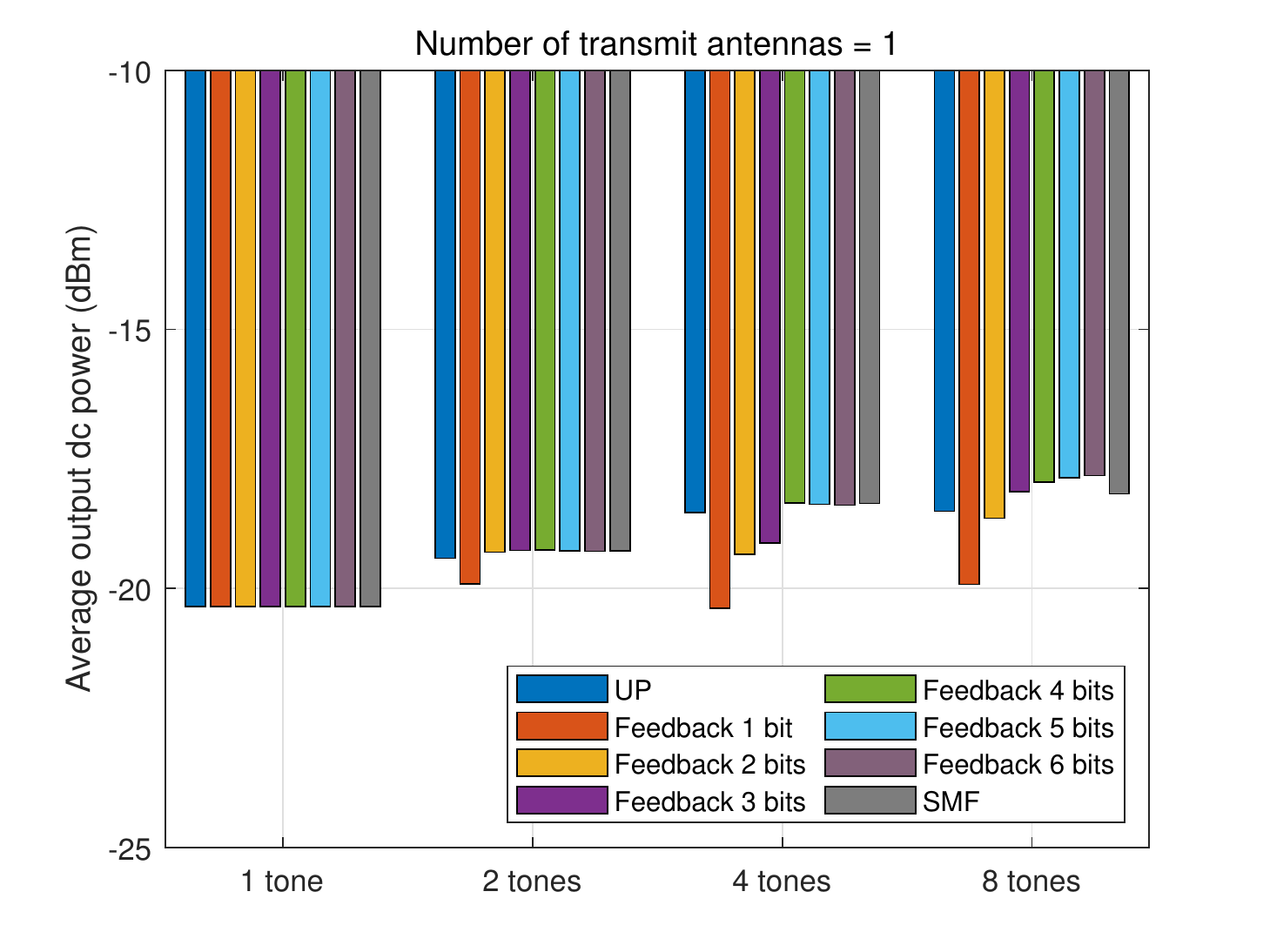}
\par\end{centering}
\caption{\label{fig:WF}Average output dc power for the three strategies versus
the number of tones with only 1 transmit antenna}
\end{figure}

Next, we show the benefit of closed-loop WPT system with only adaptive
waveform using limited feedback. To that end, we consider the three
strategies with only 1 transmit antenna and different numbers of tones,
1, 2, 4, and 8. We measure the output dc power at locations L1-L15
and average the output dc power over the 15 locations. The average
output dc power for the three strategies versus the number of tones
with only 1 transmit antenna is shown in Fig. \ref{fig:WF}. We make
the following observations.

1) The average output dc power increases with the number of tones
for the three strategies, showing the benefit of using multi-sine
waveform in WPT to increase the output dc power.

2) The closed-loop SMF and limited feedback strategies with more than
3 feedback bits achieves higher output dc power than the open-loop
strategy UP, showing the benefit of using closed-loop strategies to
adapt waveform to the wireless multipath fading channels.

3) The average output dc power for the limited feedback strategy increases
with the number of feedback bits, or the number of codewords, which
is same as the adaptive beamforming only case. When the codebook has
enough diverse codewords, the limited feedback strategy has a similar
performance to the SMF strategy.

Overall, the measurement results in Fig. \ref{fig:WF} show the benefit
of closed-loop WPT system with only adaptive waveform using limited
feedback to increase the output dc power.

\begin{figure}[t]
\begin{centering}
\includegraphics[width=8.5cm]{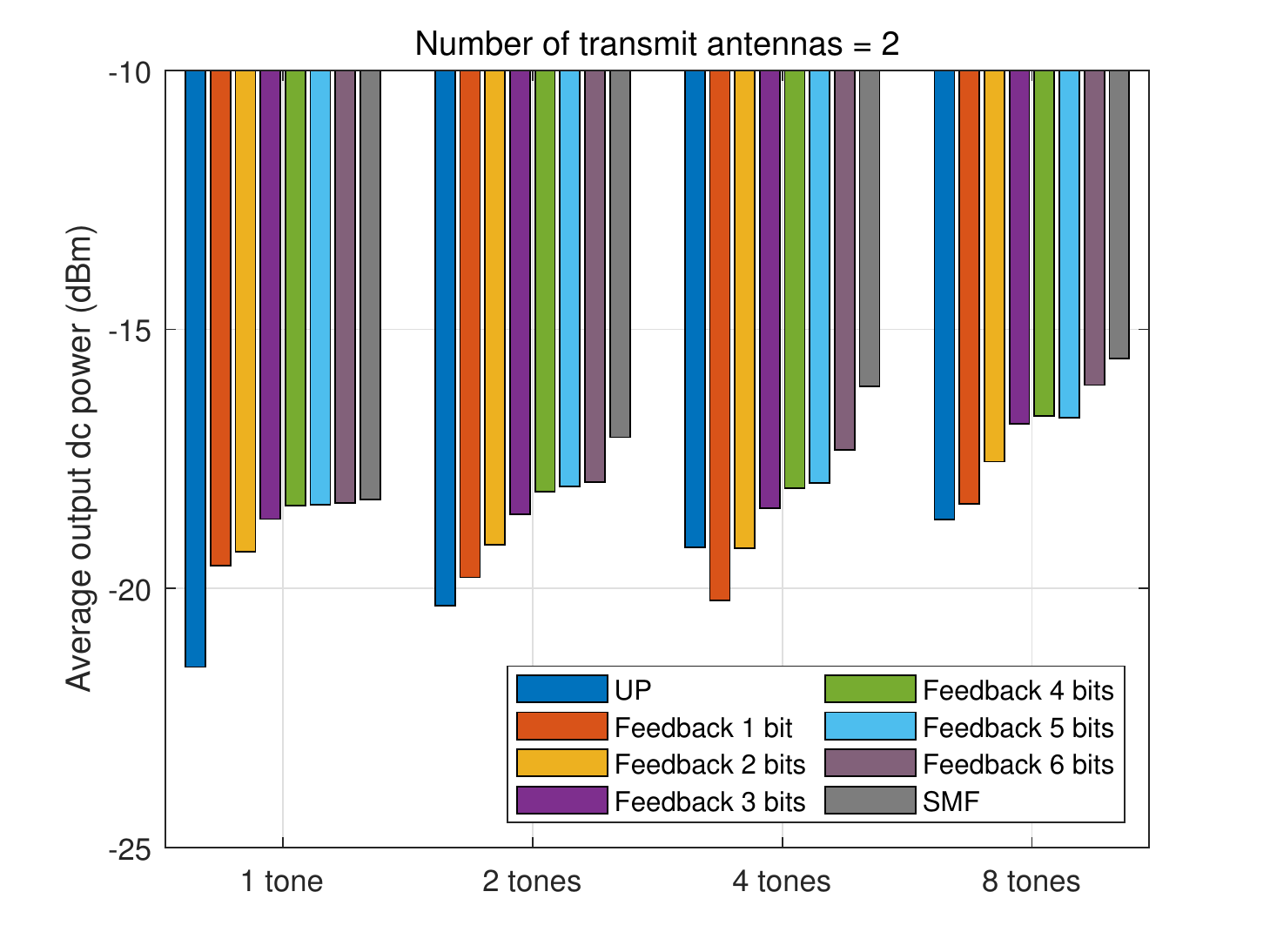}
\par\end{centering}
\begin{centering}
(a)
\par\end{centering}
\begin{centering}
\includegraphics[width=8.5cm]{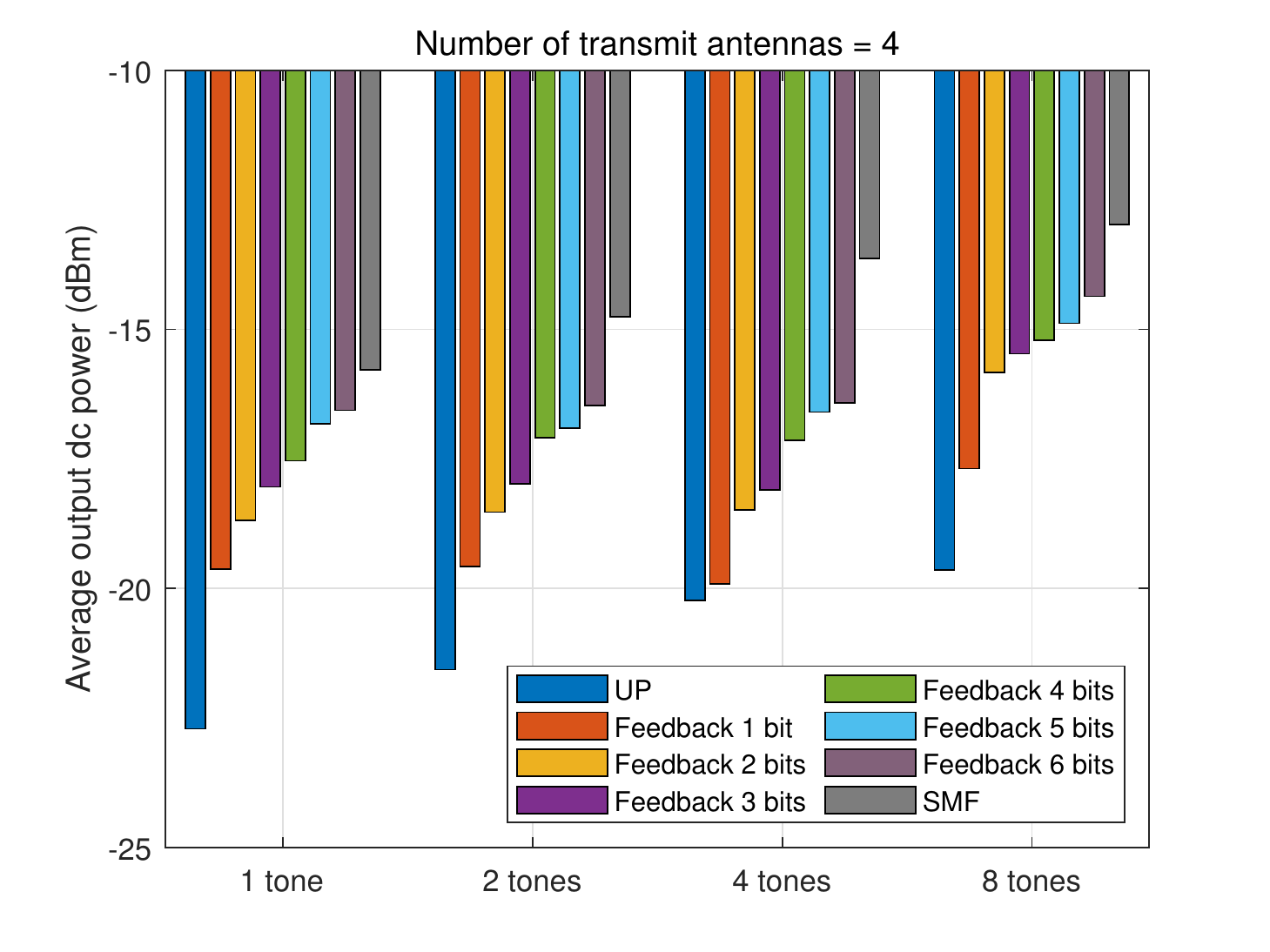}
\par\end{centering}
\begin{centering}
(b)
\par\end{centering}
\caption{\label{fig:WF-BF}Average output dc power for the three strategies
versus the number of tones with (a) 2 transmit antennas and (b) 4
transmit antennas.}
\end{figure}

\subsection{Adaptive Waveform and Beamforming}

Finally, we show the benefit of closed-loop WPT system with joint
adaptive waveform and beamforming using limited feedback. To that
end, we consider the three strategies with different numbers of transmit
antennas, 2, 4, and different numbers of tones, 1, 2, 4, and 8. We
measure the output dc power at locations L1-L15 and average the output
dc power over the 15 locations. The average output dc power for the
three strategies versus the number of tones with different numbers
of transmit antennas is shown in Fig. \ref{fig:WF-BF}. We make the
following observations.

1) Comparing the 8-tone cases and 1-tone cases in Fig. \ref{fig:WF-BF}
(a) and (b), we can find that using the joint adaptive waveform and
beamforming achieves higher output dc power than using the adaptive
beamforming only.

2) Comparing the 1-transmit-antenna case in Fig. \ref{fig:WF} and
the 2/4-transmit-antenna cases in Fig. \ref{fig:WF-BF}, we can find
that using the joint adaptive waveform and beamforming achieves higher
output dc power than using the adaptive waveform only.

3) The closed-loop SMF and limited feedback strategies with more than
1 feedback bit achieves higher output dc power than the open-loop
strategy UP, showing the benefit of using closed-loop strategies.

4) The average output dc power for the limited feedback strategy increases
with the number of feedback bits, or the number of codewords. When
the codebook has enough diverse codewords, the limited feedback strategy
has a similar performance to the SMF strategy.

\begin{figure}[t]
\begin{centering}
\includegraphics[width=8.5cm]{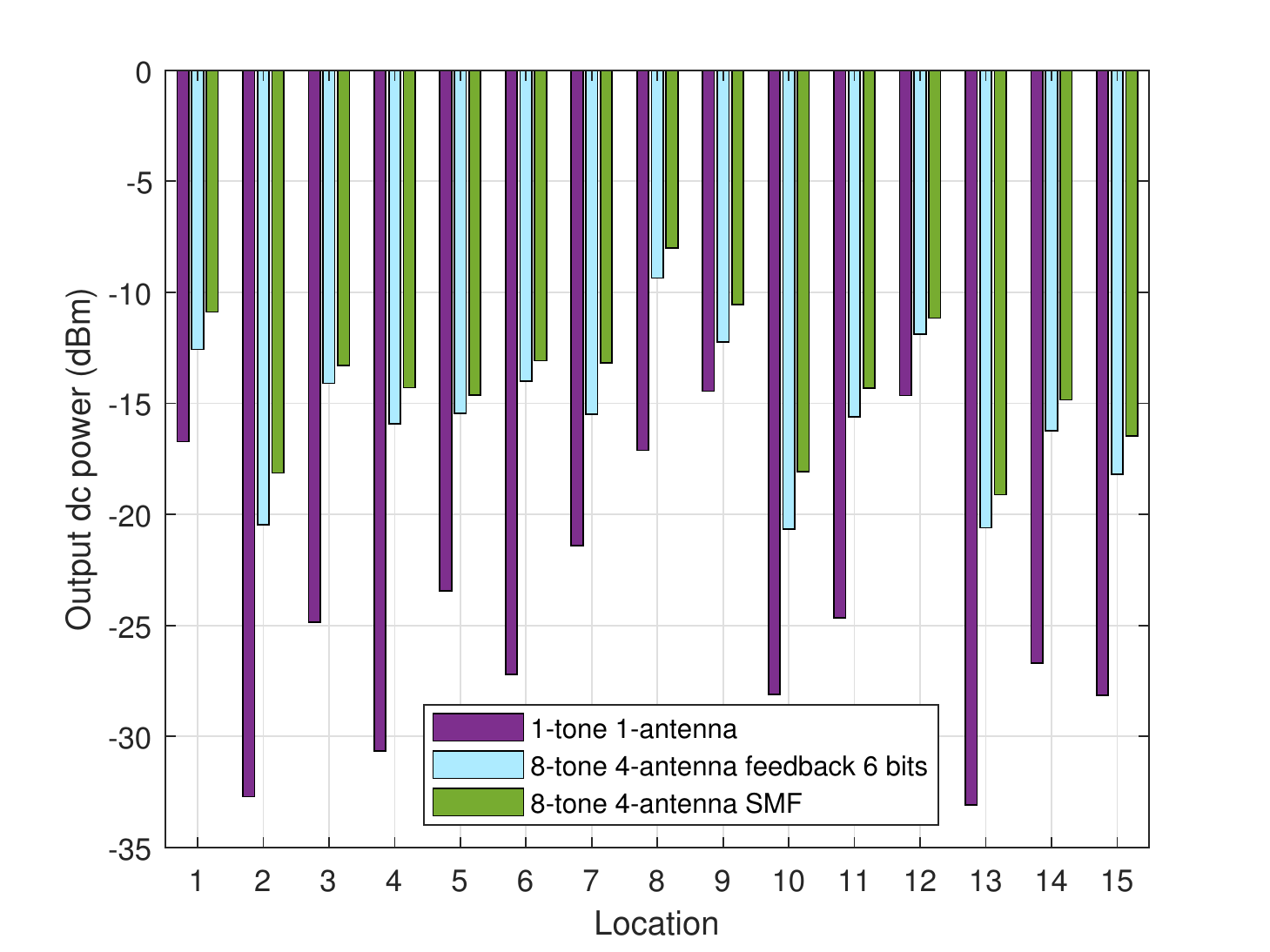}
\par\end{centering}
\caption{\label{fig:PouT}Output dc power for the 1-tone 1-antenna WPT system,
the proposed 8-tone 4-antenna closed-loop WPT system with feeding
back 6 bits, and the 8-tone 4-antenna closed-loop WPT system using
SMF at different locations.}
\end{figure}

To further show the benefit of the proposed closed-loop WPT system
with adaptive waveform and beamforming, we show the output dc power
at different locations for i) the conventional 1-tone 1-antenna WPT
system, ii) the proposed 8-tone 4-antenna closed-loop WPT system with
feeding back 6 bits, and iii) the benchmark 8-tone 4-antenna closed-loop
WPT system using SMF in Fig. \ref{fig:PouT}. Compared with the conventional
1-tone 1-antenna WPT system, using the proposed closed-loop WPT system
with adaptive waveform and beamforming can increase the output dc
power by 2.2-14.7 dB. In addition, the limited feedback strategy has
similar output dc power to the SMF strategy while having the benefit
that it does not require knowing the CSI.

Overall, the measurement results in Fig. \ref{fig:WF-BF} and Fig.
\ref{fig:PouT} show the benefit of closed-loop WPT system with adaptive
waveform and beamforming using limited feedback to increase the output
dc power without requiring the knowledge of CSI.

\section{Conclusions}

We design, prototype, and experimentally validate a closed-loop WPT
system with adaptive waveform and beamforming using limited feedback
to increase the output dc power. Spatial and frequency domains are
jointly exploited by adaptive multi-sine waveform and multi-antenna
beamforming at the transmitter in WPT system to adapt the wireless
multipath fading channel and increase the output dc power.

A closed-loop architecture for WPT based on a codebook design and
a low complexity over-the-air limited feedback using an IEEE 802.15.4
RF interface is proposed. With the designed codebook and limited feedback,
the channel estimation and accurate CSI can be avoided and more importantly
the multi-sine waveform and multi-antenna beamforming at the transmitter
can be optimized to adapt to the multipath fading channel in real
time.

The proposed closed-loop WPT with adaptive waveform and beamforming
using limited feedback is prototyped by a Software Defined Radio equipment
and measured in a real indoor environment. A closed-loop WPT system
based on cable-feedback and an open-loop WPT system are also measured
as comparison benchmarks. The measurement results show that using
closed-loop adaptive multi-sine waveform and multi-antenna beamforming
can effectively increase the output dc power. Compared with the conventional
1-tone 1-antenna WPT system, the proposed closed-loop WPT system with
adaptive waveform and beamforming can increase the output dc power
by up to 14.7 dB. In addition, compared with the cable-feedback based
closed-loop WPT system, the proposed closed-loop WPT system using
limited feedback can achieve similar performance while it is more
practical and beneficial since it does not require any computationally
complex and energy consuming channel estimation implementation at
the receiver and does not rely on knowing the CSI accurately at the
transmitter.


\begin{thebibliography}{10}
	\providecommand{\url}[1]{#1}
	\csname url@samestyle\endcsname
	\providecommand{\newblock}{\relax}
	\providecommand{\bibinfo}[2]{#2}
	\providecommand{\BIBentrySTDinterwordspacing}{\spaceskip=0pt\relax}
	\providecommand{\BIBentryALTinterwordstretchfactor}{4}
	\providecommand{\BIBentryALTinterwordspacing}{\spaceskip=\fontdimen2\font plus
		\BIBentryALTinterwordstretchfactor\fontdimen3\font minus
		\fontdimen4\font\relax}
	\providecommand{\BIBforeignlanguage}[2]{{%
			\expandafter\ifx\csname l@#1\endcsname\relax
			\typeout{** WARNING: IEEEtran.bst: No hyphenation pattern has been}%
			\typeout{** loaded for the language `#1'. Using the pattern for}%
			\typeout{** the default language instead.}%
			\else
			\language=\csname l@#1\endcsname
			\fi
			#2}}
	\providecommand{\BIBdecl}{\relax}
	\BIBdecl
	
	\bibitem{2013_MM_WPT_CUT}
	Z.~{Popovic}, ``Cut the cord: Low-power far-field wireless powering,''
	\emph{IEEE Microwave Magazine}, vol.~14, no.~2, pp. 55--62, 2013.
	
	\bibitem{ShanpuShen_2020_TIE_CSong}
	C.~Song, P.~Lu, and S.~Shen, ``Highly efficient omnidirectional integrated
	multi-band wireless energy harvesters for compact sensor nodes of
	internet-of-things,'' \emph{IEEE Trans. Ind. Electron.}, pp. 1--1, 2020.
	
	\bibitem{ShanpuShen2017_AWPL_DPTB}
	S.~Shen, C.~Y. Chiu, and R.~D. Murch, ``A dual-port triple-band {L}-probe
	microstrip patch rectenna for ambient {RF} energy harvesting,'' \emph{IEEE
		Antennas Wireless Propag. Lett.}, vol.~16, pp. 3071--3074, 2017.
	
	\bibitem{2018_TMTT_RFEH_OnPaper}
	V.~{Palazzi}, J.~{Hester}, J.~{Bito}, F.~{Alimenti}, C.~{Kalialakis},
	A.~{Collado}, P.~{Mezzanotte}, A.~{Georgiadis}, L.~{Roselli}, and M.~M.
	{Tentzeris}, ``A novel ultra-lightweight multiband rectenna on paper for {RF}
	energy harvesting in the next generation {LTE} bands,'' \emph{IEEE
		Transactions on Microwave Theory and Techniques}, vol.~66, no.~1, pp.
	366--379, 2018.
	
	\bibitem{ShanpuShen2019_TIE_HybridCombining}
	S.~{Shen}, Y.~{Zhang}, C.~{Chiu}, and R.~{Murch}, ``A triple-band high-gain
	multibeam ambient {RF} energy harvesting system utilizing hybrid combining,''
	\emph{IEEE Trans. Ind. Electron.}, vol.~67, no.~11, pp. 9215--9226, 2020.
	
	\bibitem{ShanpuShen2017_TAP_EHPIXEL}
	S.~Shen, C.~Y. Chiu, and R.~D. Murch, ``Multiport pixel rectenna for ambient
	{RF} energy harvesting,'' \emph{IEEE Trans. Antennas Propag.}, vol.~66,
	no.~2, pp. 644--656, Feb. 2018.
	
	\bibitem{2019_TMTT_EH_GridArray}
	Y.~{Hu}, S.~{Sun}, H.~{Xu}, and H.~{Sun}, ``Grid-array rectenna with wide angle
	coverage for effectively harvesting {RF} energy of low power density,''
	\emph{IEEE Transactions on Microwave Theory and Techniques}, vol.~67, no.~1,
	pp. 402--413, 2019.
	
	\bibitem{2019TAP_EH_optimalAngular}
	E.~{Vandelle} \emph{et~al.}, ``Harvesting ambient {RF} energy efficiently with
	optimal angular coverage,'' \emph{IEEE Trans. Antennas Propag.}, vol.~67,
	no.~3, pp. 1862--1873, March 2019.
	
	\bibitem{ShanpuShen2019_TMTT_Freqdepend}
	S.~Shen, Y.~Zhang, C.-Y. Chiu, and R.~Murch, ``An ambient {RF} energy harvesting system where the
	number of antenna ports is dependent on frequency,'' \emph{IEEE Trans.
		Microw. Theory Tech.}, vol.~67, no.~9, pp. 3821--3832, Sep. 2019.
	
	\bibitem{ShanpuShen_2020_IoTJ_AmRFEH}
	S.~Shen, Y.~Zhang, C.-Y. Chiu, and R.~Murch, ``Directional multiport ambient
	{RF} energy-harvesting system for the internet of things,'' \emph{IEEE
		Internet of Things Journal}, vol.~8, no.~7, pp. 5850--5865, 2021.
	
	\bibitem{2019TAP_WPT_WeiLinDrivenLoop}
	W.~{Lin} and R.~W. {Ziolkowski}, ``Electrically small huygens {CP} rectenna
	with a driven loop element maximizes its wireless power transfer
	efficiency,'' \emph{IEEE Trans. Antennas Propag.}, pp. 1--1, 2019.
	
	\bibitem{2019_TMTT_WPT_compactRectenna}
	A.~{Okba}, A.~{Takacs}, and H.~{Aubert}, ``Compact rectennas for
	ultra-low-power wireless transmission applications,'' \emph{IEEE Transactions
		on Microwave Theory and Techniques}, vol.~67, no.~5, pp. 1697--1707, 2019.
	
	\bibitem{2020_TMTT_WPT_AmRFEH_metasurface}
	L.~{Li}, X.~{Zhang}, C.~{Song}, W.~{Zhang}, T.~{Jia}, and Y.~{Huang}, ``Compact
	dual-band, wide-angle, polarization-angle-independent rectifying metasurface
	for ambient energy harvesting and wireless power transfer,'' \emph{IEEE
		Transactions on Microwave Theory and Techniques}, pp. 1--1, 2020.
	
	\bibitem{2020_Access_Meta_rectenna}
	M.~A. {Aldhaeebi} and T.~S. {Almoneef}, ``Highly efficient planar metasurface
	rectenna,'' \emph{IEEE Access}, vol.~8, pp. 214\,019--214\,029, 2020.
	
	\bibitem{2018TAP_WPT_TCA}
	T.~S. {Almoneef}, F.~{Erkmen}, M.~A. {Alotaibi}, and O.~M. {Ramahi}, ``A new
	approach to microwave rectennas using tightly coupled antennas,'' \emph{IEEE
		Trans. Antennas Propag.}, vol.~66, no.~4, pp. 1714--1724, 2018.
	
	\bibitem{2020_TMTT_RFEH_TeeShirt}
	J.~{Antonio Estrada} \emph{et~al.}, ``{RF}-harvesting tightly coupled rectenna
	array tee-shirt with greater than octave bandwidth,'' \emph{IEEE Transactions
		on Microwave Theory and Techniques}, vol.~68, no.~9, pp. 3908--3919, 2020.
	
	\bibitem{2019_TMTT_Rectifier_Single_dual_rectifier}
	M.~{Huang}, Y.~L. {Lin}, J.~{Ou}, X.~y.~{Zhang}, Q.~W. {Lin}, W.~{Che}, and
	Q.~{Xue}, ``Single- and dual-band rf rectifiers with extended input power
	range using automatic impedance transforming,'' \emph{IEEE Transactions on
		Microwave Theory and Techniques}, vol.~67, no.~5, pp. 1974--1984, 2019.
	
	\bibitem{2017_TMTT_Rectifier_UWB}
	J.~{Kimionis}, A.~{Collado}, M.~M. {Tentzeris}, and A.~{Georgiadis}, ``Octave
	and decade printed {UWB} rectifiers based on nonuniform transmission lines
	for energy harvesting,'' \emph{IEEE Transactions on Microwave Theory and
		Techniques}, vol.~65, no.~11, pp. 4326--4334, 2017.
	
	\bibitem{2019_TMTT_EH_Rectifier_BoosterRegulator}
	S.~{Fan}, Z.~{Yuan}, W.~{Gou}, Y.~{Zhao}, C.~{Song}, Y.~{Huang}, J.~{Zhou}, and
	L.~{Geng}, ``A 2.45-{GHz} rectifier-booster regulator with impedance matching
	converters for wireless energy harvesting,'' \emph{IEEE Transactions on
		Microwave Theory and Techniques}, vol.~67, no.~9, pp. 3833--3843, 2019.
	
	\bibitem{2019_TMTT_Rectifier_Insensitive}
	S.~N. {Daskalakis}, A.~{Georgiadis}, G.~{Goussetis}, and M.~M. {Tentzeris}, ``A
	rectifier circuit insensitive to the angle of incidence of incoming waves
	based on a wilkinson power combiner,'' \emph{IEEE Transactions on Microwave
		Theory and Techniques}, vol.~67, no.~7, pp. 3210--3218, 2019.
	
	\bibitem{2020_TMTT_Rectifier_CoupleTransmission}
	F.~{Zhao}, D.~{Inserra}, G.~{Gao}, Y.~{Huang}, J.~{Li}, and G.~{Wen},
	``High-efficiency microwave rectifier with coupled transmission line for
	low-power energy harvesting and wireless power transmission,'' \emph{IEEE
		Transactions on Microwave Theory and Techniques}, pp. 1--1, 2020.
	
	\bibitem{2020_TMTT_Rectifier_Efficient}
	S.~A. {Rotenberg}, S.~K. {Podilchak}, P.~D.~H. {Re}, C.~{Mateo-Segura},
	G.~{Goussetis}, and J.~{Lee}, ``Efficient rectifier for wireless power
	transmission systems,'' \emph{IEEE Transactions on Microwave Theory and
		Techniques}, vol.~68, no.~5, pp. 1921--1932, 2020.
	
	\bibitem{ShanpuShen2019_JSSC_Reconfigure}
	Z.~Zeng, S.~Shen, X.~Zhong, X.~Li, C.-Y. Tsui, A.~Bermak, R.~Murch, and
	E.~Sánchez-Sinencio, ``Design of sub-gigahertz reconfigurable {RF} energy
	harvester from -22 to 4 {dBm} with 99.8$\%$ peak {MPPT} power efficiency,''
	\emph{IEEE J. Solid-State Circuits}, vol.~54, no.~9, pp. 2601--2613, Sep.
	2019.
	
	\bibitem{2017_TMTT_RFEH_Solar}
	J.~{Bito}, R.~{Bahr}, J.~G. {Hester}, S.~A. {Nauroze}, A.~{Georgiadis}, and
	M.~M. {Tentzeris}, ``A novel solar and electromagnetic energy harvesting
	system with a 3-{D} printed package for energy efficient internet-of-things
	wireless sensors,'' \emph{IEEE Transactions on Microwave Theory and
		Techniques}, vol.~65, no.~5, pp. 1831--1842, 2017.
	
	\bibitem{ShanpuShen2019_TMTT_RF_Solar}
	Y.~{Zhang}, S.~{Shen}, C.~Y. {Chiu}, and R.~{Murch}, ``Hybrid {RF}-solar energy
	harvesting systems utilizing transparent multiport micromeshed antennas,''
	\emph{IEEE Transactions on Microwave Theory and Techniques}, vol.~67, no.~11,
	pp. 4534--4546, 2019.
	
	\bibitem{2017_TOC_WPT_YZeng_Bruno_RZhang}
	Y.~{Zeng}, B.~{Clerckx}, and R.~{Zhang}, ``Communications and signals design
	for wireless power transmission,'' \emph{IEEE Trans. Commun.}, vol.~65,
	no.~5, pp. 2264--2290, May 2017.
	
	\bibitem{clerckx2021wireless}
	B.~Clerckx, K.~Huang, L.~R. Varshney, S.~Ulukus, and M.-S. Alouini, ``Wireless
	power transfer for future networks: Signal processing, machine learning,
	computing, and sensing,'' \emph{ArXiv}, 2101.04810, 2021.
	
	\bibitem{2015_MM_WPT_waveform}
	A.~{Boaventura}, D.~{Belo}, R.~{Fernandes}, A.~{Collado}, A.~{Georgiadis}, and
	N.~B. {Carvalho}, ``Boosting the efficiency: Unconventional waveform design
	for efficient wireless power transfer,'' \emph{IEEE Microwave Magazine},
	vol.~16, no.~3, pp. 87--96, 2015.
	
	\bibitem{2014TMTT_WPT_MutliSine_Spatial_Power_Combination}
	A.~J.~S. Boaventura \emph{et~al.}, ``Spatial power combining of multi-{s}ine
	signals for wireless power transmission applications,'' \emph{IEEE Trans.
		Microw. Theory Techn.}, vol.~62, no.~4, pp. 1022--1030, Apr. 2014.
	
	\bibitem{2016_TMTT_RFEH_multitone}
	F.~{Bolos}, J.~{Blanco}, A.~{Collado}, and A.~{Georgiadis}, ``Rf energy
	harvesting from multi-tone and digitally modulated signals,'' \emph{IEEE
		Transactions on Microwave Theory and Techniques}, vol.~64, no.~6, pp.
	1918--1927, 2016.
	
	\bibitem{2017_TMTT_RFID_Multisine}
	A.~J.~S. {Boaventura} and N.~B. {Carvalho}, ``The design of a high-performance
	multisine {RFID} reader,'' \emph{IEEE Transactions on Microwave Theory and
		Techniques}, vol.~65, no.~9, pp. 3389--3400, 2017.
	
	\bibitem{2017_TMTT_WPT_INVIVO}
	Z.~{Liu}, Z.~{Zhong}, and Y.~{Guo}, ``In vivo high-efficiency wireless power
	transfer with multisine excitation,'' \emph{IEEE Transactions on Microwave
		Theory and Techniques}, vol.~65, no.~9, pp. 3530--3540, 2017.
	
	\bibitem{2014_MWCL_WPT_Optimal_Waveform}
	A.~{Collado} and A.~{Georgiadis}, ``Optimal waveforms for efficient wireless
	power transmission,'' \emph{IEEE Microw. Wireless Compon. Lett.}, vol.~24,
	no.~5, pp. 354--356, May 2014.
	
	\bibitem{2018_MM_WPT_Bruno_1GWPT}
	B.~{Clerckx}, A.~{Costanzo}, A.~{Georgiadis}, and N.~{Borges Carvalho},
	``Toward {1G} mobile power networks: {RF}, signal, and system designs to make
	smart objects autonomous,'' \emph{IEEE Microw. Mag.}, vol.~19, no.~6, pp.
	69--82, Sep. 2018.
	
	\bibitem{ShanpuShen_2020_TWC_MIMO_WPT}
	S.~Shen and B.~Clerckx, ``Beamforming optimization for {MIMO} wireless power
	transfer with nonlinear energy harvesting: {RF} combining versus {DC}
	combining,'' \emph{IEEE Trans. Wireless Commun.}, vol.~20, no.~1, pp.
	199--213, 2021.
	
	\bibitem{2019_TMTT_WPT_Selective_Tracking}
	D.~{Belo}, D.~C. {Ribeiro}, P.~{Pinho}, and N.~{Borges Carvalho}, ``A
	selective, tracking, and power adaptive far-field wireless power transfer
	system,'' \emph{IEEE Transactions on Microwave Theory and Techniques},
	vol.~67, no.~9, pp. 3856--3866, 2019.
	
	\bibitem{2020_TMTT_WPT_PhasedArray}
	B.~{Yang}, X.~{Chen}, J.~{Chu}, T.~{Mitani}, and N.~{Shinohara}, ``A 5.8-{GHz}
	phased array system using power-variable phase-controlled magnetrons for
	wireless power transfer,'' \emph{IEEE Transactions on Microwave Theory and
		Techniques}, vol.~68, no.~11, pp. 4951--4959, 2020.
	
	\bibitem{2016_TMTT_WPT_TimeArray}
	D.~{Masotti}, A.~{Costanzo}, M.~{Del Prete}, and V.~{Rizzoli},
	``Time-modulation of linear arrays for real-time reconfigurable wireless
	power transmission,'' \emph{IEEE Transactions on Microwave Theory and
		Techniques}, vol.~64, no.~2, pp. 331--342, 2016.
	
	\bibitem{2016_MM_WPT_SMART}
	A.~{Costanzo} and D.~{Masotti}, ``Smart solutions in smart spaces: Getting the
	most from far-field wireless power transfer,'' \emph{IEEE Microwave
		Magazine}, vol.~17, no.~5, pp. 30--45, 2016.
	
	\bibitem{2017_IEEEACCESS_WPT_BlindBF}
	P.~S. {Yedavalli}, T.~{Riihonen}, X.~{Wang}, and J.~M. {Rabaey}, ``Far-field
	{RF} wireless power transfer with blind adaptive beamforming for internet of
	things devices,'' \emph{IEEE Access}, vol.~5, pp. 1743--1752, 2017.
	
	\bibitem{2017_TWC_WPT_Prototyping_Zolerta}
	K.~W. {Choi}, L.~{Ginting}, P.~A. {Rosyady}, A.~A. {Aziz}, and D.~I. {Kim},
	``Wireless-powered sensor networks: How to realize,'' \emph{IEEE Trans.
		Wireless Commun.}, vol.~16, no.~1, pp. 221--234, Jan 2017.
	
	\bibitem{2018_TSP_WPT_Prototyping_RSSI}
	S.~{Abeywickrama}, T.~{Samarasinghe}, C.~K. {Ho}, and C.~{Yuen}, ``Wireless
	energy beamforming using received signal strength indicator feedback,''
	\emph{IEEE Trans. Signal Process.}, vol.~66, no.~1, pp. 224--235, Jan 2018.
	
	\bibitem{2018_MWCL_WPT_3rdHarmonic}
	H.~{Zhang}, Y.~{Guo}, S.~{Gao}, and W.~{Wu}, ``Wireless power transfer antenna
	alignment using third harmonic,'' \emph{IEEE Microwave and Wireless
		Components Letters}, vol.~28, no.~6, pp. 536--538, 2018.
	
	\bibitem{2019_MWCL_WPT_3rdHarmonic2}
	H.~{Zhang}, Y.~{Guo}, S.~{Gao}, Z.~{Zhong}, and W.~{Wu}, ``Exploiting third
	harmonic of differential charge pump for wireless power transfer antenna
	alignment,'' \emph{IEEE Microwave and Wireless Components Letters}, vol.~29,
	no.~1, pp. 71--73, 2019.
	
	\bibitem{2020_TMTT_WPT_SecondHarmonic}
	S.~D. {Joseph}, Y.~{Huang}, S.~S.~H. {Hsu}, A.~{Alieldin}, and C.~{Song},
	``Second harmonic exploitation for high-efficiency wireless power transfer
	using duplexing rectenna,'' \emph{IEEE Transactions on Microwave Theory and
		Techniques}, pp. 1--1, 2020.
	
	\bibitem{2016_TSP_WPT_Bruno_Waveform}
	B.~{Clerckx} and E.~{Bayguzina}, ``Waveform design for wireless power
	transfer,'' \emph{IEEE Trans. Signal Process.}, vol.~64, no.~23, pp.
	6313--6328, Dec 2016.
	
	\bibitem{2017_TSP_WPT_Bruno_Yang_Large}
	Y.~{Huang} and B.~{Clerckx}, ``Large-scale multiantenna multisine wireless
	power transfer,'' \emph{IEEE Trans. Signal Process.}, vol.~65, no.~21, pp.
	5812--5827, Nov 2017.
	
	\bibitem{2017_AWPL_WPT_Bruno_LowComplexity}
	B.~{Clerckx} and E.~{Bayguzina}, ``Low-complexity adaptive multisine waveform
	design for wireless power transfer,'' \emph{IEEE Antennas Wireless Propag.
		Lett.}, vol.~16, pp. 2207--2210, 2017.
	
	\bibitem{2018_TWC_WPT_Bruno_HYang_LimitedFeedback}
	Y.~{Huang} and B.~{Clerckx}, ``Waveform design for wireless power transfer with
	limited feedback,'' \emph{IEEE Trans. Wireless Commun.}, vol.~17, no.~1, pp.
	415--429, Jan 2018.
	
	\bibitem{shen2020joint}
	S.~Shen and B.~Clerckx, ``Joint waveform and beamforming optimization for
	{MIMO} wireless power transfer,'' \emph{IEEE Trans. Commun.}, pp. 1--1, 2021.
	
	\bibitem{2019_Junghoon_Prototyping}
	J.~Kim, B.~Clerckx, and P.~D. Mitcheson, ``Signal and system design for
	wireless power transfer: Prototype, experiment and validation,'' \emph{IEEE
		Trans. Wireless Commun.}, vol.~19, no.~11, pp. 7453--7469, 2020.
	
	\bibitem{2021_WCL_WPT_RangeExpansion}
	J.~Kim and B.~Clerckx, ``Range expansion for wireless power transfer using
	joint beamforming and waveform architecture: An experimental study in indoor
	environment,'' \emph{IEEE Wireless Communications Letters}, pp. 1--1, 2021.
	
	\bibitem{ShanpuShen_2020_TIE_WPT_DAS}
	S.~{Shen}, J.~{Kim}, C.~{Song}, and B.~{Clerckx}, ``Wireless power transfer
	with distributed antennas: System design, prototype, and experiments,''
	\emph{IEEE Trans. Ind. Electron.}, pp. 1--1, 2020.
	
\end{thebibliography}
\end{document}